\documentstyle[12pt,aasms4,psfig]{article}
\def\gsimeq
{\hbox{\raise0.5ex\hbox{$>\lower1.06ex\hbox{$\kern-1.07em{\sim}$}$}}}
\def\lsimeq
{\hbox{\raise0.5ex\hbox{$<\lower1.06ex\hbox{$\kern-1.07em{\sim}$}$}}}
\def\pn{\par\noindent}

\begin{document}

\title{The Complex X-ray Spectrum of 3C 273: {\it ASCA} Observations}

\author{M. Cappi$^{1,2}$, M. Matsuoka$^1$, C. Otani$^1$ \& K.M. Leighly$^{1,3}$}

\affil{ $^1$ The Institute of Physical and Chemical Research (RIKEN), \\ Hirosawa 2-1, Wako,
Saitama 351-01, Japan}

\affil{ $^2$ Istituto per le Tecnologie e Studio Radiazioni Extraterrestri (ITeSRE), CNR, \\
Via Gobetti 101, I-40129 Bologna, Italy}

\affil{ $^3$ Columbia Astrophysics Laboratory, Columbia University, \\ 538 West 120th Street, 
New York, NY 10027}

\centerline{Accepted for publication in P.A.S.J.}

\begin{abstract}
Results obtained from 9 X-ray observations of 3C 273 performed by {\it ASCA} are 
presented (for a total exposure time of about 160 000 s).
The analysis and interpretation of the results is complicated by the fact
that 4 of these observations were used for on-board calibration of the CCDs spectral
response. In particular, we had to pay special attention to the low energy band and 5--6
keV energy range where systematic effects could distort a correct interpretation of
the data.

The present standard analysis shows that, in agreement with official recommendations,
a conservative systematic error (at low energies) of $\sim$ 2--3 $\times$
10$^{20}$ cm$^{-2}$ must be assumed when analyzing {\it ASCA} SIS data.
A soft-excess, with variable flux and/or shape, has been clearly detected
as well as flux and spectral variability
that confirm previous findings with other observatories. 
An anti-correlation is found between the spectral index and the flux in the 
2-10 keV energy range. 
With the old response matrices, an iron emission line feature with EW $\sim$
50--100 eV was initially
detected at $\sim$ 5.6-5.7 keV ($\sim$ 6.5-6.6 keV in the quasar frame) 
in 6 observations and, in two occasions, the line was resolved ($\sigma \sim$ 0.2-0.6 keV).
Comparison with the Crab spectrum indicates however that this feature was mostly
due to remaining calibration uncertainties between 5--6 keV.
Indeed, fitting the data with the latest publicly available calibration matrices,
we find that the line remains unambiguously significant in (only) the two observations
with lowest fluxes where it is weak (EW $\sim$ 20-30 eV), narrow and consistent
with being produced by Fe K$_{\alpha}$ emission from neutral matter.

Overall, the observations are qualitatively consistent with a variable, non-thermal 
X-ray continuum emission, i.e., a power law with $\Gamma$ $\sim$ 1.6 (possibly produced in 
the innermost regions of the radio-optical jet), plus underlying ``Seyfert-like''
features, i.e., a soft-excess and Fe K$_{\alpha}$ line emission.
The data are consistent with some contribution (up to a few 10\% level in the {\it ASCA}
energy band) from a
``Seyfert-like'' direct continuum emission, i.e. a power law
with $\Gamma$ $\sim$ 1.9 plus a reflection component, as well.
When the continuum (jet) emission is in a low state, the spectral features produced by the 
Seyfert-like spectrum (soft-excess, iron line and possibly a steep power law plus a reflection
continuum) are more easily seen.

\end{abstract}


\section{Introduction}

The remarkable discovery by EGRET on-board {\it CGRO} that blazars (i.e. BL Lacertae 
objects and flat-spectrum radio quasars) are strong $\gamma$-ray emitters has 
drawn in recent years the attention of the astronomical community to this 
class of objects.
Observations indicate that the overall energy distribution of blazars 
shows the signature of two different types of emission mechanisms: beamed 
non-thermal jet radiation producing the overall broad band 
(from radio to $\gamma$-rays) continuum emission common to all blazars, 
and quasi-isotropic thermal radiation by an accretion-disk
producing a ``UV Bump'' observed in a large number of quasars and Seyfert galaxies
but absent in BL Lac objects (e.g. Sambruna, Maraschi \& Urry, 1996, Elvis et al. 1994).
The non-thermal continuum consists of IR-optical and $\gamma$-ray peaks.
The first peak is interpreted in terms of synchrotron emission 
and the second peak in terms of inverse Compton emission (see Urry \& Padovani
1995 for a review on the subject).
From object to object, the observed different spectral characteristics 
may be due to the relative importance of one emission mechanism 
to the other which, in turn, is likely to be related to the amount 
of beaming in one object or the other (e.g. Dondi \& Ghisellini 1995). 

One of the most well-studied and characteristic example of blazars is the 
bright quasar 3C 273 (z$\simeq$0.158). It is a good example where both non-thermal and
thermal emission mechanisms 
might be observed because its broad band energy distribution exhibits 
two large peaks, one peaking in the IR-optical and one peaking 
in the $\gamma$-rays with a `UV bump'' superimposed on it (Courvoisier et al. 1987,
Lichti et al. 1995, von Montigny et al. 1997).
The study of its X-ray properties may provide important clues
to understand the origin of both emission mechanisms because a) soft-X-ray excess emission 
has been observed and interpreted as the high-energy tail of the UV bump 
(Turner et al. 1985, Courvoisier et al. 1987, Walter et al. 1994,
Leach, Mc Hardy \& Papadakis 1995)
and b) the 2-10 keV spectrum which is most likely associated to the $\gamma$-ray 
emission is known to be variable in time and shape (Turner et al. 1985), thus allowing for tests
of different X- and $\gamma$-ray emission models.
\pn
To clarify the mechanism responsible for the X-ray emission, high quality data are
first necessary to disentangle the contributions from the different spectral
components, namely the jet and Seyfert components.

In this paper, we report on observations with {\it ASCA} during the first year 
of the mission.
The source spectral properties are shown with particular attention to calibration 
uncertainties, most relevant in this source because it was used for on-board calibration
of the CCDs. 
We show evidence of complex spectral features (soft-excess, Fe K emission line, flux and 
spectral correlated variability) and discuss their possible interpretation. Throughout the 
analysis we use $H_{0}$ = 50 km s$^{-1}$ Mpc$^{-1}$ and $q_{0}$ = 0.

\section{Observations and Data Reduction}

Over the period from 1993 June to 1993 December, 3C 273 was observed 9 times with
the gas imaging spectrometer (GIS) and solid-state spectrometer (SIS) on-board the
{\it ASCA} satellite (Tanaka, Inoue \& Holt 1994).
The observation log is given in Table 1. The SIS was
operating in 1 CCD Faint mode during all the observations.
Observations 2, 3, 5 and 6 have been used to calibrate each chip of the SIS
(Dotani et al. 1996) and the remaining pointings are part of a multi-wavelength campaign
on 3C 273 (von Montigny et al., 1997).
Dark frame error (DFE) and echo corrections (Otani \& Dotani 1994)
have been applied to all the data.
After removing hot and flickering pixels, standard selection criteria were used to
select good observational intervals.
The most relevant were an elevation and bright Earth angles greater than 5$^{\circ}$
and 25$^{\circ}$ respectively, and a magnetic cutoff rigidity greater than
8 GeV/c for SIS and 7 GeV/c for GIS. For all observations, source counts were
extracted using circular region centered on the source of radius 6$^{\prime}$ for the
GIS and 3$^{\prime}$ for the SIS. Hard X-ray emission was detected with {\it SIGMA}
from a location about 15$^{\prime}$ away from 3C 273 (at R.A.[1950]=12$^h$27$^m$20$^s$,
DEC[1950]=02$^\circ$30$^\prime$ with an associated error of $\sim$ 5$^\prime$) (Jourdain
et al. 1992). We find no trace of such source in the {\it ASCA} field of views.
Background spectra were obtained from the edges of the
chips in the SIS and from the blank sky files in the GIS. The GIS background region was
always chosen in regions uncontaminated by NGC 6552 and at similar off-axis distance
as the source region (see Appendix A of Cappi et al. 1997 for a detailed
discussion).
Owing to the high count-rate of this bright source and the fact that the background
typically contributed less than a few percent to the average count-rates in each
observations, different choices of backgrounds (e.g. blank sky backgrounds for SIS
or local backgrounds for GIS) did not introduce significant differences in the spectral
results reported below. 
In total, {\it ASCA} collected about half a million counts per detector for an
effective exposure time of $\sim$ 160 Ks for GIS and $\sim$ 130 Ks for SIS.
Data preparation and spectral analysis have been done using version 1.2 of the
XSELECT package and version 9.0 of the XSPEC program (Arnaud et al 1991).
Official EA/PSF-V1.0 files, gis[23]v4\_0.rmf and rsp1.1alphaPI1.6 matrices (released 
in 1994) were used for all the observations reported in the following. Newer XRT
calibration files (EA/PSF-V2.0) and related GIS and SIS responses were also used in \S 3.4 for
the study of the iron line emission.

\section{Results}

GIS2/3 and SIS0/1 pulse-height spectra were binned so as to have at least 200 counts
per bin, in the energy ranges 0.7-10 keV and 0.4-10 keV, respectively.
Light curves were accumulated for all observing periods, but none of these indicated
significant variability (although variations as high as $\sim$ 10-20\% per observation 
period cannot be excluded).
All photons were, thus, accumulated observation-by-observation for the spectral analysis.
It is emphasized that 3C 273 was used to calibrate the SIS response function
imposing consistency with the GIS results (GIS was assumed to be well calibrated by fitting
the Crab data with a single absorbed power law model; see Dotani et al. 1996 for details).
Therefore, great care must be taken when interpreting the SIS results since systematic
effects could be present.
As a rule, {\it absolute} values obtained from the SIS (only) could not, in principle,
be trusted. However, {\it relative} measurements (like flux and/or spectral variability)
obtained from comparing different observations are likely to give valuable information
since, in this case, systematic effects should cancel each other.
Moreover, all the observations were taken during the first year of the mission,
and the last 8 observations were performed within a period of 12 days. Thus, we 
assumed that the instrument did not change significantly from one observation 
to the other.

\subsection{Flux Variability}

First, GIS and SIS data were fitted separately with a single absorbed power law model,
with the column density, $N_{\rm H}$, free to vary.
Throughout the analysis, the column density is calculated in the 
observer frame (i.e. z=0).
This model gives an acceptable description of all the spectra. The best-fitting
model results are given in Table 2.
It is emphasized that the 2-10 keV flux measurements
(column 4 in Table 2) do not vary significantly even if $N_{\rm H}$ is fixed at the Galactic
value (\S 3.3) or if an iron emission line is added to the model (\S 3.4).
Figure 1 clearly shows that the flux varied in time by up to
$\sim$ 60\% on a time-scale of $\sim$ 200 days. Significant (20--30\%) shorter-term
variations on a time-scale of $\sim$ 1 day are also evident in the light-curve.
The luminosity varied, accordingly, from L$_{(2-10 \rm keV)}$ $\sim$ 1.4 $\times$ 
10$^{46}$ erg s$^{-1}$ in observation 1 to a L$_{(2-10 \rm keV)}$ $\sim$ 2.1 
$\times$ 10$^{46}$ erg s$^{-1}$ in observation 9.
Note that the photon indices obtained from the GIS and SIS are in excellent agreement,
while there are slight, but significant, discrepancies in the measurement of the
low energy absorption column. This effect will be considered in more detail in the
following section.
The 2--10 keV flux measured with the SIS is also systematically $\sim$ 15--20\% higher
than that obtained from the GIS. The discrepancy is currently 
attributed to cross calibration uncertainties that have been largely reduced
to less than 10\% with the
introduction of new (EA/PSF-V2.0) XRT calibration files.
With the matrices used in the present analysis, the SIS flux can be considered to be
the most reliable with an absolute uncertainty of less than $\sim$ 10\%
(K. Arnaud et al., {\it ASCA} calibration uncertainties,
http:$//$heasarc.gsfc.nasa.gov$/$docs$/$asca$/$cal$\_$probs.html). 

The absolute value and long time-scale variations of the 2-10 keV flux of 3C 273 are
roughly consistent with previous observations with {\it EXOSAT} and {\it Ginga} satellites
(Turner et al. 1990). However, the better quality and temporal
sampling of the present observations also reveal for the first time
day-to-day variations of the 2--10 keV X-ray flux as large as $\sim$ 20\%.

\subsection{Soft Excess and Excess Absorption}

90\% confidence contours for the column densities versus photon indices
obtained from the SIS spectra with the above absorbed power law model (Table 2) are shown in
Fig. 2.
The vertical lines represent the Galactic absorption (full line) and associated estimated
errors (dotted lines) as obtained from measurements at 21 cm (Dickey \& Lockman 1990).
Fig. 2 clearly illustrates three interesting results: (1) The best-fit column density
was significantly lower than the Galactic value during the first observation. This result can
be promptly interpreted as evidence for a soft-excess in the data; (2) in all other
observations, absorption was significantly, and systematically, higher than the Galactic
value; (3) small, but significant, variations in the photon index are measured from
observation to observation.

Note that, though the GIS is less sensitive than the SIS at low energies, the soft-excess
is also detected in the GIS data. Indeed, the GIS spectrum gives an upper
limit for $N_{\rm H}$ of $\sim$ 0.65 $\times$ 10$^{20}$ cm$^{-2}$ during the first
observation (significantly lower than $N_{\rm Hgal}$ $\sim$ 1.79 $\times$ 10$^{20}$
cm$^{-2}$) while all other observations give $N_{\rm H}$ values consistent with the
Galactic absorption.
Although this measurement is in agreement with that reported by
Yaqoob et al. (1994), these authors did not report the detection of a soft-excess
for this observation probably because, in their work, SIS spectra were consistent with
the Galactic  absorption. This discrepancy is likely to be attributed to the fact that
Yaqoob et al. (1994) used older response matrices obtained from the preliminary
calibrations of the GIS and SIS instruments.

The evidence for a soft-excess during observation 1 is further strengthened by the
ratio of the pulse height spectra of observation 1 divided by
observation 9 of the summed SIS0 and SIS1 spectra (upper panel)
and the summed GIS2 and GIS3 spectra (lower panel) shown in Fig. 3.
Similar results were obtained dividing observation 1 with the other observations, 
thus excluding the possibility that the steepening of the ratios below about 2 keV could be 
due to a slight increase of the absorption measured during observation 9.
Since pulse height spectral ratios directly compare the raw
count-rates from two datasets, problems related to the calibration uncertainties of the
instrumental response are in principle removed. These further confirm the
presence of a continuum soft-excess emission at energies below $\sim$ 1.5 keV. 
In order to parameterize the measured soft-excess, two-component emission models
were fitted to the GIS and SIS data of observation 1 with the absorption fixed at 
the Galactic value.
Normalizations from the two detectors were free to vary independently.
Results from the model-fittings are given in Table 3.
Attempts to model the soft-excess with a Raymond \& Smith (1977) plasma model
and a warm absorber model, as proposed in previous work, are not reported here since
both models gave poor fits to the data, consistent with the absence of any blend of
emission lines around 0.8--0.9 keV (quasar frame) or absorption features around
0.7--1.0 keV (quasar frame).
Acceptable fits (Table 3) were obtained from either a black-body (with kT $\sim$ 100 eV)
plus power law model, a bremsstrahlung (with kT $\sim$ 240 eV) plus power law model
or a double power law model (with $\Gamma_{\rm soft} \sim$ 3).

It is interesting to note that the improvement in $\chi^2$ obtained with a
double power law model is statistically significant when compared to the
black-body ($\Delta \chi^2 = 10$) or bremsstrahlung ($\Delta \chi^2 = 5$) models at
$\sim$ 99.8\% and $\sim$ 96\% confidence levels, respectively. 
This is similar to the findings of Leach et al. (1995) obtained from high quality
{\it ROSAT} PSPC data and confirm that the soft-excess is better modeled by a
power law than by a thermal emission model. This issue will be further addressed in
\S 4.2.
Moreover, it is likely that this soft-excess is intrinsically variable with time
(in shape or in intensity) because our simulations indicate that, if constant, it
should have been detected at least during observation 7, when the source was only
$\sim$ 15\% brighter than in observation 1.

The following 4 calibration observations and 4 AO-I observations are, instead, all
consistent with a constant absorption column of $\sim$ 3--4 $\times$ 10$^{20}$ cm$^{-2}$.
This value is significantly higher than the Galactic absorption by
$\sim$ 2-3 $\times$ 10$^{20}$ cm$^{-2}$. However, a physical interpretation of this
result cannot be made because the SIS responses were calibrated from some of these
observations (Table 1) assuming that 3C 273 was absorbed by the Galactic column only.
Instead, this column should, by definition, be
taken as a systematic error of the {\it ASCA} SIS response function for a standard
analysis like the one presented in this work. Similar results have been
reported from independent work on 3C 273 (Hayashida et al. 1995) and the Coma cluster
(Dotani et al. 1996; Hashimotodani, private communication).

\subsection{Photon Index Variations}

Figure 2 also suggests small but significant variations of the power law
photon index, at least from observation 2 to observation 9.
However, because of the presence of the soft-excess and extra-absorption
mentioned above, these variations may be due to an incorrect modeling 
of the low energy absorption.
To avoid such complications, only the data above 2 keV were considered.
Since the photon indices obtained by separately fitting
the GIS and SIS data were all consistent at $\sim$ 90\% confidence level, the data
from the two instruments were fitted simultaneously tying the fitting parameters
together but allowing the relative normalizations to be free.
A single power law model plus absorption fixed at the
Galactic value was used in all fits.
It is pointed out that since the SIS best-fit spectra normally require a value of $N_{\rm H}$
systematically higher than the Galactic value when fitted between $\sim$ 0.4--10 keV
(see \S 3.2), by fixing the absorption at the Galactic value, a systematic
error will be introduced on the {\it absolute} value of the SIS photon indices.
However, this (small) effect will not affect the measurements of
{\it relative} values presented below.
GIS results are not affected by these uncertainties. 
Best-fitting results are given in Table 4. The variations of the 2--10 keV
photon indices with the observed 2--10 keV fluxes are shown in Fig. 4 for the
(a) GIS+SIS data (all observations) and (b) GIS+SIS
data, excluding calibration observations n. 2, 3 and 5. The reason we ignored
observations 2, 3 and 5 in the dataset (b) is that these were performed
using chips 0, 2 and 3 of SIS0 and chips 0, 1, 2 of
SIS1 in which the gain (relative to the standard chip n.1 of SIS0 and chip n. 3 of SIS1)
are uncertain by as much as $\sim$ 2\% (Dotani et al. 1996). This could
introduce systematic errors in the spectral slope of as much as $\sim$ 0.05
which are much larger than our statistical errors and could, therefore, distort any 
significant statistical correlation.

Figure 4 shows evidence for a spectral index change in 3C 273.
Though the variations are not
large ($\Delta \Gamma \lsimeq$ 0.1), they are significant in both datasets (Table 5).
This point is further illustrated in Figure 3 which shows that, though the data are somewhat
noisy above $\sim$ 8 keV, the pulse height spectrum between 2-10 keV of observation 1 is
steeper than that of observation 9, for both the GIS and SIS data.
Table 5 shows the results from $\chi^2$-tests against constancy, linear correlation
coefficients and rank correlation coefficients for the two datasets considered.
The most interesting result obtained from this analysis is that there is
a trend (most noticeable in Fig. 4) for the 2--10 keV photon index to be
anti-correlated with the flux, i.e. a spectral flattening as the source brightens.
This result is significant at a $\sim$ 99.95\%
and $\sim$ 81\% confidence level for the dataset (a) and at a $\sim$ 99.999\% and
$\sim$ 89\% confidence level for dataset (b), for a linear and non-parametric correlation
respectively (Table 5). It is emphasized that the same trend was observed also
in both SIS and GIS instruments taken individually, though with somewhat lower significance.
However, it should be noted that the significance of this result depends mostly 
on the 2 data points from observations 1 and 9.
If real, this behavior is opposite to that commonly seen in Seyfert galaxies
(e.g. Mushotzky, Done \& Pounds 1993 and references therein) and, on the contrary,
provides an interesting analogy with BL Lac objects as discussed in \S 4.


\subsection{Iron Emission Line}

The spectra have been inspected for the presence of an iron emission line.
Our first analysis clearly reveals the presence of a broad
($\sigma \sim$ 0.4 keV) Fe K line with EW $\sim$ 90--100 eV during
observations 1 and 7 (Table 6). Confidence contours in the Fe K parameters space
$\sigma$--E and EW--E (observer frame) obtained from fitting the observation 1
data are shown in Fig. 5 as an example. A narrow line is also significant in 4 out
of 7 of the remaining observations.
The emission lines are all consistent with a neutral or mildly ionized
iron emission line (at 6.4 keV in the quasar's frame) at
$\sim$ 90\% confidence level. We find no significant
correlation between the equivalent width of the Fe K line and the computed X-ray
flux but it is interesting that the 2 strongest lines are detected while 3C 273 was in its
lowest states (i.e. observations 1 and 7 reported in Table 6).

However, as is evident from the dashed contours shown in Fig. 5, the GIS spectra of
the Crab which were used to calibrate the GIS
exhibit a miscalibration at similar energies which can be
modeled with a broad ($\sigma \simeq$ 0.4 $\pm$ 0.05 keV) line at E $\simeq$ 5.7 
$\pm$ 0.05 keV with an EW $\simeq$ 50 $\pm$ 15 eV (see also Ebisawa et al. 1996 for
more details about this miscalibration in the Crab's GIS spectra).

New calibration matrices (ascaarf v2.62 with version 2.0 of the XRT matrices) have,
however, recently been distributed to the {\it ASCA} guest observers
through the {\it ASCA} Goddard Guest Observer Facility (K. Ebisawa,
http:$//$heasarc.gsfc.nasa.gov$/$docs$/$asca$/ $xrt$\_$new$\_$response$\_$announce$/$announce.html). The new calibrations reduce the systematic differences in flux between
GIS and SIS (see \S 3.1) and corrects the GIS and SIS response matrices for
the $\sim$ 5--6 keV spectral feature detected in the Crab spectra
(Fukazawa, Ishida \& Ebisawa 1997, but see also Gendreau \& Yaqoob 1997).
It is emphasized that the use of this improved calibration does not affect our
conclusions on flux
variability, soft-excess and spectral variability because, as stated in \S 3, only
relative measurements have been considered.
It gives, however, significantly different results on the iron emission line measurements.
Now, spectral fitting yields narrow ($\sigma$ $<$ 0.21 keV and $<$ 0.6 keV)
and weaker, EW ($\sim$ 25 eV and 27 eV, observer frame) lines during observations
1 and 7, respectively (see Table 7).
The line is also reduced in all other observations yielding only upper limits
ranging between $\sim$ 10--30 eV.
Analysis with the latest available calibration suggests, therefore, that the line
detected during observations 1 and 7 is very likely to be real but that, contrary
to our first results with previous calibration, it is consistent with being narrow and
with having an EW $\sim$ 20-30 eV (observer frame).
In the quasar's frame, the Fe K line best-fit parameters are
E$\simeq$6.47$^{+0.09}_{-0.07}$ keV, 6.31$^{+0.06}_{-0.06}$ keV and
EW$\simeq$29$^{+23}_{-15}$ eV, 31$^{+18}_{-18}$ eV for observations 1 and 7, respectively.
These results agree with the emission of a fluorescent narrow Fe K$_{\alpha}$ line from
neutral matter (Makishima 1986).



\section{Discussion}

\subsection{On the X-ray Continuum Emission: Signatures of a Jet Component ?}

The {\it ASCA} observations have shown that the 2-10 keV flux is variable with 
time by up to $\sim$ 60\% on a time-scale of $\sim$ 200 days and shows day-to-day 
variations as large as $\sim$ 20\%.
Thus, unless it is relativistically beamed, the hard X-ray source must 
be smaller than R = c $\times$ $\Delta$t $\sim$ 10$^{16}$cm where the source
doubling time-scale, i.e. the time necessary for the source to vary by a factor
of two, can roughly be estimated as $\Delta$t $\sim$
${F_{\rm init} \over{\Delta F}}$$\times$${\Delta t_{\rm obs}\over{1+z}}$, where z is 
the source redshift.
It is well known that 3C 273 shows a one-sided jet which is clearly observed at
both radio and optical
wavelengths (Bahcall et al. 1995 and references therein) which is most likely 
produced by synchrotron emission from relativistic electrons (3C 273 
exhibits superluminal motion with $\beta_{\rm app} \simeq$ 8.0 $\pm$ 1.0, e.g. Vermeulen 
\& Cohen 1994).
The main jet consists of a number of bright, elongated, knots extending from $\sim$ 30
kpc to $\sim$ 50 kpc from the quasar with kpc-scale transverse dimensions 
in both the radio and optical images. Therefore, the dimensions of the radio-optical jet are
about 6 orders of magnitude larger than the X-ray source dimension.
This, together with the high X-ray luminosity of $\sim$ 1.5-2 $\times$ 10$^{46}$ 
erg s$^{-1}$ inferred from the X-ray data, excludes the possibility that the bulk of the 
X-ray emission is produced by the jet's knots only (whatever the emission 
mechanism is).
Therefore, if it is related to the jet, the 2-10 keV X-ray emitting region must be
located in the innermost regions of the jet.
These conclusions are consistent with high resolution X-ray imaging (e.g.
{\it ROSAT} HRI) of the jet emission of 3C 273 (R\"oser et al. 1996).

For the first time in this source, we find evidence for 
a statistically significant anti-correlation
between the 2-10 keV flux 
and spectral index, implying hardening of the spectrum as the source brightens.
This is opposite to lower luminosity AGNs (e.g. 
Seyfert 1 galaxies) which often show a positive correlation (Grandi et al. 1992,
Mushotzky, Done \& Pounds 1993, and references therein) but similar to the 
anti-correlation flux-spectral index often observed in BL Lacertae objects 
(at least in X-ray selected BL Lac objects; e.g. Giommi et al. 1990, 
Urry et al. 1996).
This suggests that the bulk of the X-ray emission in 3C 273 is produced by a 
mechanism similar to the one supposed in BL Lac objects, i.e. a process 
related to the existence of a beam of relativistic particles (e.g. Fichtel 1995).
This is consistent with the recent results of von Montigny et al. (1997) which show 
that the multi-wavelength spectrum (from radio to $\gamma$-rays) of 3C 273 can be 
explained by any of the most prominent theoretical models (e.g. synchrotron 
self-Compton (SSC, Maraschi, Ghisellini \& Celotti 1992), inverse Compton on 
external photons from an accretion disk or a broad-line region (EC, Sikora, Begelman \& 
Rees 1994) or synchrotron 
from ultra-relativistic electrons and positrons in a proton-induced cascade (Mannheim \& 
Biermann 1992)) 
for the explanation of the high $\gamma$-ray emission of BL Lac objects.
On the basis of the present {\it ASCA} data, it is not possible, however, to distinguish 
between the different theoretical models since all of these could explain the 
observed spectral variability. Synchrotron losses, for example, have often been 
invoked to explain a steepening with decreasing flux (e.g. in PKS 2155-304, 
Sembay et al. 1993; in H0323+022, Kohmura et al. 1994; in Mkn 421, Takahashi 
et al. 1996). In these cases, the X-ray spectra were interpreted as the high energy
tail of synchrotron emission, while the X-ray spectrum of 3C 273 is more 
likely due to Compton emission (von Montigny et al. 1997). However, one could 
expect that a similar behavior may hold also in the ``Compton'' bump, though
perhaps at a different time-scale because of a possibly different emission region.
Moreover, hysteresis ``clockwise'' flux-index
relations were reported in these BL Lac objects. Further observations are clearly
necessary to test this hypothesis in 3C 273.

Alternatively, there might be a mixture of, say, SSC and EC contributing to the 
X-ray emission, the spectrum being harder when the EC flux increases.
Or else, as discussed next, there might be a contribution from a Seyfert-like
spectrum with a steeper spectral slope, e.g. $\Gamma$ $\sim$ 1.7 as commonly
seen in Seyfert galaxies (Mushotzky, Done \& Pounds 1993), which steepens the spectrum
as the source becomes fainter. 
A small contribution of $\sim$10-20\% in the 
2-10 keV band could steepen the observed spectrum by $\Delta \Gamma$ $\sim$ 0.1 thus 
explain the flux-index anti-correlation.

\subsection{On the X-ray Spectral Features: Signatures of a Seyfert-like Component ?}

Another interesting result from our analysis is that the spectra intermittently
show clear evidence for a separate soft component below $\sim$ 2 keV. 
This component was observed only during the first observation, when the source was 
faintest.
The best fit values for this component (see \S 3.2) are in good agreement with previous 
{\it EXOSAT}, {\it EINSTEIN},{\it GINGA} and {\it ROSAT} findings (e.g. Courvoisier 
et al. 1987, Wilkes \& Elvis 1987, Turner et al. 1990, Staubert 1992, Leach et al. 1995), in particular 
if one allows for cross-calibration uncertainties and considers that previous 
observations usually found, or fixed, the photon index between 2-10 keV at a 
slightly flatter value of $\Gamma \sim$ 1.5 than found with {\it ASCA}.
The issue of interpreting the soft continuum cannot, however, be addressed
in more detail with the present {\it ASCA} data since, as explained in $\S 3$,
absolute values obtained from the SIS cannot in principle be trusted because
3C 273 was used to calibrate the SIS response (see Dotani et al. 1996 for details on the 
calibration procedure).
For example, because of the systematic excess absorption found in \S 3.2 
which is most likely attributed to a calibration uncertainty, we cannot
trust absolute values of the best-fit parameters 
for the soft-excess.
Only relative measurements can be considered, which tell us 
that a soft excess component is indeed required by the data during 
observation 1 and not during the following observations.

It should be pointed out, however, that as found by Leach et al. (1995) from the
{\it ROSAT} PSPC data, the soft component is modeled better by a single power law with 
absorption at the Galactic value ($\Delta \chi^{2}$ = 10 and 5 compared to the 
black-body and bremsstrahlung models, respectively).
This may indicate that the data are not well described by a model with a concave shape
but rather prefer a straight or convex model.
At this point, it is interesting to note that a recent {\it Beppo-SAX}
observation of 3C 273 performed in July 1996, when the source was at very low flux level
($\sim$ 7.1 $\times$ 10$^{-11}$ erg cm$^{-2}$ s$^{-1}$), shows evidence for both
a soft excess emission (below $\sim$ 0.3 keV) and a strong absorption structure
at $\sim$ 0.5 keV indicating, possibly, the presence of a warm absorber in 3C 273
(Grandi et al. 1997). {\it ASCA} observed 3C 273 simultaneously with {\it Beppo-SAX}
on that occasion. Results of this observation will be presented elsewhere
(Yaqoob et al., in preparation).
Fitting the {\it ASCA} observation 1 with a warm absorber model only does not
however improve the spectral fitting significantly since, clearly, the data require
some extra emission below $\sim$ 2 keV that cannot be explained with absorption
features alone. Addition of an absorption edge at $\sim$ 0.5 keV to the double power 
model increases the quality of the fit slightly but not significantly.


The {\it ASCA} spectra also reveal evidence for iron line emission in 3C 273 
on two different occasions, during observations 1 and 7, when the source 
was faintest. The best-fit parameters of the Fe K line obtained from the fitting of the 
SIS data only with the latest available response matrices are consistent 
with the line being narrow and weak (EW $\sim$ 20-30 eV in both observations 
1 and 7). These values are in reasonable agreement with the {\it GINGA} results 
of Turner et al. (1990) and the {\it Beppo-SAX} results of Grandi et al. (1997), who
both found a significant iron line in 3C 273 when the source was in a low 
(F$_{(2-10 \rm keV)}$ $\lsimeq$ 1 $\times$ 10$^{-10}$ erg cm$^{-2}$ s$^{-1}$) state,
similar to the results reported here.
The line emission and, also the soft excess emission suggest that, at
 least during observations 1 and 7, a Seyfert-like component underlying a
dominant simple power-law (jet ?) component contributes significantly to the X-ray
spectrum of 3C 273.
In order to give a rough quantitative estimate of the contribution from the
Seyfert and to test this overall picture,
we tried to fit the broad-band ($\sim$0.4--300 keV) {\it ASCA} plus {\it OSSE} spectrum
with a model consisting in the sum of: a power-law with $\Gamma$ $\simeq$ 1.5-1.6, 
a ``typical'' Seyfert-like spectrum (Nandra \& Pounds 1994), i.e. a power-law (with $\Gamma$=1.9) plus a
reflection component (with R=${\rm normalization\ of\ the\ reflected\ component\over{normalization\ of\ the\ direct\ component}}$=1 corresponding to a 2$\pi$ coverage of the
reflector) and
associated iron line (with an equivalent width fixed at 150 eV, with respect to the Seyfert
continuum, e.g. George \& Fabian 1991), and a black-body.
The {\it ASCA} spectra are from observation 1 and the {\it OSSE} spectrum is from the
observation performed during 1991 June 15-28 (viewing period n.3 reported in
Johnson et al. 1995). 
As demonstrated by the unfolded spectrum and residuals shown in Fig. 6, an
acceptable fit is obtained with a black-body temperature of kT $\sim$ 100 eV, slightly
reduced compared to the results given in Table 3 because of the Seyfert
power law contribution, and with the Seyfert-like spectrum contributing 
about 30\% at 1 keV and about 10-20\% in the 2-10 keV energy band to the
total spectrum.
However, it is stressed that while the present data are consistent with this picture,
there is actually no direct evidence for a steep ``Seyfert-like'' power-law, since
its contribution cannot be unambiguously disentangled from the softer black-body
component. However, we have also shown in \S 3.2 that the soft-excess emission
is likely to be variable in time and, as discussed above, is better fitted with a model
with a convex shape rather than a concave one. These may be the indication that some
direct, steep, Seyfert-like component contributes partly to the soft-excess emission
of 3C 273.

An alternative explanation for the Fe K line emission may be reflection
off optically thick matter (e.g. broad-line region blobs and/or a molecular torus and/or 
an accretion disk, located near the jet region) from the jet continuum
itself. In such case, the estimation of the expected iron line intensity is complicated
by the possibility that the X-ray and hard X-ray radiation could be beamed in the direction
of, or away from, the reflecting matter and its precise evaluation is beyond the scope of 
this paper. The effect of the unknown effective covering factor
of the reflector should also be taken into account.

\section{Conclusions}

To date, {\it ASCA} has observed 3C 273 10 times.
Results from the first 9 observations, all performed during the first year of
the mission have been presented here. These confirm and expand the evidence that the
X-ray emission of 3C 273 is complex, with different spectral components contributing
to its X-ray emission. Because 4 of the 9 observations were used for the on-board
calibration of the CCDs, great care had to be taken when interpreting the
observational results for this source. As a rule, {\it absolute} values, in
particular those obtained from the SIS, require a detailed estimate and knowledge of the
instrumental systematic errors to be trusted.
{\it Relative} measurements (like flux and/or spectral variability) obtained from
comparing different observations are, however, more reliable since they should not be
affected by calibration uncertainties.

With this caveat in mind, it is found that:
\pn
\begin{enumerate}
\item
A conservative systematic error at low energies that corresponds to an extra-absorption
column of $\sim$ 2--3 $\times$ 10$^{20}$ cm$^{-2}$ is found for the SIS response, consistent
with the ASCA Team's official prescriptions.
\item
2--10 keV flux variations by up to $\sim$ 60\% on a time-scale of $\sim$ 200 days and
day-to-day variations as large as $\sim$ 20\% were observed.
\item
Extra soft X-ray emission is required by the data during the first observation, when the
source was in its lowest flux level.
\item
Flux and spectral slope variations are clearly detected as well and, for the first time, there
is a statistically significant evidence that the index and flux are anti-correlated.
\item
Iron line emission is detected in (only) the two observations with the lowest flux levels.
The line is in both cases weak (EW$\sim$20-30 eV), but statistically significant at more than 99\% confidence level, narrow and consistent with Fe K$_{\alpha}$ emission from
neutral matter.
\end{enumerate}

We then speculate that all the above observable properties of the 
X-ray spectrum of 3C 273 can be interpreted in terms of the sum of two 
emission mechanisms. These are a non-thermal emission from the 
innermost regions of the jet which dominates the 2-10 keV region and whose signatures 
are the spectral variability and a flat ($\Gamma \sim 1.6$) power law continuum
(that extrapolates well into higher energies), plus a diluted Seyfert-like spectrum
whose signatures are the soft-excess and iron line emission.
The newly discovered index-flux anti-correlation may be interpreted either by intrinsic
variations of the jet power law index or by some contribution of the Seyfert-like
continuum spectrum (say, a power law with $\Gamma \sim 1.9$) as the jet component varies.
The overall scenario predicts that when the (dominant) jet component is in a low
flux state, the spectral features produced by the Seyfert-like spectrum should be more
easily detected (owing for variability of the Seyfert-like spectrum itself).

\section*{ACKNOWLEDGEMENTS}

We are grateful to the {\it ASCA} team in ISAS for their operation of the satellite and to
the {\it ASCA} GOF at NASA/GSFC for their assistance in data analysis. M.C. acknowledges
colleagues in the Institute of Physical and Chemical Research (RIKEN) for their warm
hospitality and the Italian Space Agency (ASI) for financial support.
C.O. acknowledges the Special Postdoctoral Researchers Program of RIKEN for support.
KML gratefully acknowledges support through NAG5-3307 ({\it ASCA}).
We thank T. Yaqoob, T. Dotani, K. Gendreau and T. Kotani for very helpful
discussion on the calibration-related issues and H. Kubo and G. Ghisellini for usefull 
comments.



\clearpage

\begin{table}
\caption{Observations log}
\begin{center}
\begin{tabular}{ccccccccc}
\hline
\hline
\multicolumn{1}{c}{Obs.} &
\multicolumn{1}{c}{Start Time (UT)} &
\multicolumn{2}{c}{Exposure$^*$ (s)} &
\multicolumn{2}{c}{CR$^*$(s$^{-1}$)} &
\multicolumn{1}{c}{SIS Mode} &
\multicolumn{1}{c}{Comment}\\
\multicolumn{1}{c}{} &
\multicolumn{1}{c}{} &
\multicolumn{1}{c}{GIS} &
\multicolumn{1}{c}{SIS} &
\multicolumn{1}{c}{GIS} &
\multicolumn{1}{c}{SIS} &
\multicolumn{1}{c}{} &
\multicolumn{1}{c}{} &
\multicolumn{1}{c}{} \\
\hline
1 & 08/06/93 20:11 & 33300  & 24945 & 2.25 & 3.25 & 1CCD & PV (Yaqoob et al. 94)\\
2 & 15/12/93 12:34 & 18130  & 14207 & 2.85 & 4.10 & 1CCD & PV calib s0c2\\
3 & 15/12/93 23:53 & 21230  & 17033 & 3.88 & 4.96 & 1CCD & PV calib s0c0\\
4 & 16/12/93 12:11 & 13295  & 9488  & 3.67 & 4.44 & 1CCD & AOI \\
5 & 19/12/93 23:45 & 19867  & 16136 & 3.04 & 4.17 & 1CCD & PV calib s0c3\\
6 & 20/12/93 10:45 & 19390  & 15497 & 2.88 & 5.32 & 1CCD & PV calib s0c1 \\
7 & 20/12/93 22:03 & 12291  & 10398 & 2.75 & 3.69 & 1CCD & AOI \\
8 & 23/12/93 23:35 & 11605  & 10325 & 2.84 & 3.72 & 1CCD & AOI \\
9 & 27/12/93 13:58 & 10493  & 7169  & 3.39 & 4.37 & 1CCD & AOI \\
\hline
\hline
\end{tabular}
\end{center}
$^{*}$ Exposure and count-rates are average for GIS and SIS detectors.
\end{table}

\begin{table}
\caption{Results Using a Single Absorbed Power Law Model}
\begin{center}
\begin{tabular}{cccccc}
\hline
\hline
\multicolumn{1}{c}{Obs.} &
\multicolumn{1}{c}{Instrument} &
\multicolumn{1}{c}{$N_{\rm H}^a$} &
\multicolumn{1}{c}{$\Gamma$} &
\multicolumn{1}{c}{F$_{\rm X}^b$(2-10 keV)} &
\multicolumn{1}{c}{$\chi^{2}_{red}/$d.o.f.} \\
\hline
1 & GIS & $<$ 0.65 & 1.63$_{-0.01}^{+0.01}$& 1.04 & 1.34/518\\
 & SIS & $<$ 0.19 & 1.64$_{-0.01}^{+0.01}$ & 1.23 & 1.62/350\\
 & & & & & \\
2 & GIS & $<$ 1.86 & 1.62$_{-0.02}^{+0.02}$ & 1.33 & 1.04/412 \\
 & SIS  & 4.31$_{-0.59}^{+0.57}$ & 1.60$_{-0.02}^{+0.01}$ & 1.53 & 1.62/296\\
 & & & & & \\
3 & GIS & $<$ 1.80 & 1.60$_{-0.02}^{+0.01}$& 1.54 & 1.05/546\\
 & SIS & 4.48$_{-0.56}^{+0.52}$ & 1.60$_{-0.01}^{+0.02}$ & 1.70 & 1.73/358\\
 & & & & & \\
4 & GIS & 3.09$_{-1.84}^{+1.83}$ & 1.62$_{-0.02}^{+0.03}$& 1.56 & 0.95/382 \\
 & SIS & 4.18$_{-0.68}^{+0.64}$ & 1.59$_{-0.02}^{+0.02}$ & 1.71 & 1.02/445\\
 & & & & & \\
5 & GIS & $<$ 1.50 & 1.53$_{-0.01}^{+0.02}$ & 1.45 & 1.16/463\\
 & SIS & 3.54$_{-0.55}^{+0.54}$ & 1.53$_{-0.02}^{+0.02}$ & 1.58 & 1.94/319\\
 & & & & & \\
6 & GIS & 2.69$_{-1.65}^{+1.73}$ & 1.58$_{-0.02}^{+0.02}$ & 1.31 & 1.105/443\\
 & SIS & 3.67$_{-0.59}^{+0.60}$ & 1.56$_{-0.02}^{+0.02}$ & 1.51 & 1.30/300\\
 & & & & & \\
7 & GIS & $<$ 2.44 & 1.59$_{-0.02}^{+0.03}$ & 1.22 & 0.96/263\\
 & SIS & 3.55$_{-0.71}^{+0.68}$ & 1.58$_{-0.03}^{+0.02}$ & 1.44& 1.22/430\\
 & & & & & \\
8 & GIS & $<$ 3.61 & 1.56$_{-0.02}^{+0.03}$ & 1.30 & 1.09/259\\
 & SIS & 3.89$_{-0.71}^{+0.70}$ & 1.55$_{-0.02}^{+0.02}$  & 1.48 & 1.08/435\\
 & & & & & \\
9 & GIS & 3.57$_{-2.10}^{+2.15}$ & 1.54$_{-0.02}^{+0.03}$ & 1.57 & 1.01/272\\
 & SIS & 5.01$_{-0.85}^{+0.83}$ & 1.52$_{-0.03}^{+0.02}$  & 1.82 & 1.04/391\\
 & & & & & \\
\hline
\hline
\end{tabular}
\end{center}
\pn
\pn
$^{a}$ Absorption column density in units of $10^{20}$ cm$^{-2}$.
\pn
$^{b}$ Observed flux in units of $10^{-10}$ erg cm$^{-2}$ s$^{-1}$.
\pn
Note: Errors are 90 \% confidence for 2 interesting parameters ($\Delta \chi^2$=4.61).
\end{table}

\begin{table}
\caption{SIS -- Two Component Models Fitted to the Observation 1 Data}
\begin{center}
\begin{tabular}{ccccc}
\hline
\hline
\multicolumn{1}{c}{} &
\multicolumn{1}{c}{$kT/\Gamma_{\rm soft}$} &
\multicolumn{1}{c}{$\Gamma/\Gamma_{\rm hard}$} &
\multicolumn{1}{c}{L$^{a}_{\rm X}$(0.1-2.0 keV)} &
\multicolumn{1}{c}{$\chi^{2}_{red}$/d.o.f.} \\
\multicolumn{1}{c}{} &
\multicolumn{1}{c}{(eV)} &
\multicolumn{1}{c}{} &
\multicolumn{1}{c}{($10^{45}$ erg s$^{-1}$)} &
\multicolumn{1}{c}{} \\
\hline
black body + power law & $119^{+16}_{-16}$ & ${1.64^{+0.02}_{-0.02}}$ & 1.07 &1.46/869 \\
bremss. + power law & ${281^{+89}_{-70}}$&${1.64^{+1.66}_{-1.62}}$ & 2.38 &1.45/869 \\
Two power laws& ${2.99^{+0.84}_{-0.85}}$ &${1.59^{+0.08}_{-0.09}}$ & 5.70 &1.44/869 \\
\hline
\hline
\end{tabular}
\end{center}
\pn
$^{a}$ calculated from only the soft component and the SIS normalization. Values with the
GIS normalization were approximately 15\% lower.
\pn
\pn
Note: Intervals are at 90 \% confidence for 2 interesting parameters.
\end{table}

\begin{table}
\caption{Fits between 2-10 keV with a Single Absorbed Power Law Model - $N_{\rm H} \equiv N_{\rm Hgal}$}
\begin{center}
\begin{tabular}{cccc}
\hline
\hline
\multicolumn{1}{c}{Obs.} &
\multicolumn{1}{c}{$\Gamma$} &
\multicolumn{1}{c}{F$_{\rm X}^a$(2-10 keV)} &
\multicolumn{1}{c}{$\chi^{2}_{red}/$d.o.f.} \\
\hline
1  &  1.62$_{-0.01}^{+0.01}$ & 1.22 & 1.50/454\\
 & &  & \\
2 &  1.58$_{-0.02}^{+0.02}$ & 1.48 & 1.47/334\\
& &  & \\
3 &  1.59$_{-0.01}^{+0.01}$ & 1.67  & 1.40/498\\
& &  & \\
4 &  1.56$_{-0.02}^{+0.02}$ &1.69 & 1.05/434\\
& &  & \\
5 &  1.52$_{-0.01}^{+0.01}$ &  1.56& 1.22/403\\
& &  & \\
6 &  1.57$_{-0.01}^{+0.01}$ &1.49 & 1.15/371\\
& &  & \\
7 &  1.57$_{-0.02}^{+0.02}$ & 1.41& 1.24/370\\
& &  & \\
8 &  1.55$_{-0.02}^{+0.02}$ & 1.47& 1.02/378\\
& & & \\
9 &  1.52$_{-0.02}^{+0.02}$ & 1.77 & 1.06/355\\
\hline
\hline
\end{tabular}
\end{center}
\pn
\pn
$^{a}$ Observed SIS flux in units of $10^{-10}$ erg cm$^{-2}$ s$^{-1}$. GIS flux
was typically $\sim$15-20\% lower.
\pn
Note: Intervals are at 90 \% confidence for 1 interesting parameter ($\Delta \chi^2$=2.71).
\end{table}

\begin{table}
\caption{$\Gamma_{\rm 2-10 keV}$ -- F$_{\rm X}$(2-10 keV) correlations}
\begin{center}
\begin{tabular}{cccccccc}
\hline
\hline
\multicolumn{1}{c}{Instrument} &
\multicolumn{1}{c}{d.o.f.} &
\multicolumn{1}{c}{$\chi^{2,\ a}$} &
\multicolumn{1}{c}{p$^a$} &
\multicolumn{1}{c}{r$^b$} &
\multicolumn{1}{c}{p$^b$} &
\multicolumn{1}{c}{r$_{\rm s}^c$} &
\multicolumn{1}{c}{p$^c$} \\
\hline
GIS+SIS (all)& 9 & 35.1 & 5.7 $\times$ 10$^{-5}$ & $-0.64$ & $\sim$ 0.05 & -0.48 & 0.19 \\
GIS+SIS$^d$ & 6 & 20.4 & 2.3 $\times$ 10$^{-3}$ & $-0.89$ & $\sim$ 1
$\times$ 10$^{-3}$ & $-0.71$ & 0.11 \\
\hline
\hline
\end{tabular}
\end{center}
\pn
$^{a}$ $\chi^2$ value and corresponding probability (p) for a $\chi^2$ test against
constancy.
\pn
$^{b}$ Linear correlation coefficient (r) and corresponding probability (p) for a
linear correlation of the data.
\pn
$^{c}$ Spearman rank-order coefficient (r$_{\rm s}$) and corresponding probability (p) for
a non-parametric correlation of the data.
\pn
$^{d}$ GIS plus SIS data excludind observations 2, 3 and 5 which were
used to calibrate the ``non-standard'' SIS chips.
\end{table}

\begin{table}
\caption{Iron emission line best-fit parameters$^*$}
\begin{center}
\begin{tabular}{cccccc}
\hline
\hline
\multicolumn{1}{c}{} &
\multicolumn{1}{c}{$\Gamma_{\rm 2-10 keV}$} &
\multicolumn{1}{c}{E } &
\multicolumn{1}{c}{$\sigma$} &
\multicolumn{1}{c}{EW} &
\multicolumn{1}{c}{$\chi^{2}_{red}$/d.o.f.} \\
\multicolumn{1}{c}{} &
\multicolumn{1}{c}{} &
\multicolumn{1}{c}{(keV)} &
\multicolumn{1}{c}{(keV)} &
\multicolumn{1}{c}{(eV)} &
\multicolumn{1}{c}{} \\
\hline
\multicolumn{6}{c}{Old calibration software and matrices} \\
\hline
Obs. 1 & 1.63$^{+0.01}_{-0.01}$ & 5.59$^{+0.04}_{-0.04}$&0 (fixed)&40$^{+10}_{-15}$&1.44/454\\
\hfill& 1.63$^{+0.01}_{-0.01}$ & 5.67$^{+0.10}_{-0.12}$&0.43$^{+0.23}_{-0.19}$&100$^{+48}_{-34}$&1.43/453\\
\hline
Obs. 7 & 1.57$^{+0.02}_{-0.02}$ & 5.46$^{+0.04}_{-0.04}$&0 (fixed)&38$^{+17}_{-17}$&1.21/372\\
\hfill& 1.57$^{+0.02}_{-0.02}$ & 5.44$^{+0.16}_{-0.15}$&0.28$^{+0.25}_{-0.10}$&81$^{+29}_{-29}$&1.20/371\\
\hline
\hline
\multicolumn{6}{c}{New calibration software and matrices} \\
\hline
Obs. 1 & 1.56$^{+0.01}_{-0.01}$ & 5.59$^{+0.08}_{-0.06}$&0 (fixed)&25$^{+20}_{-13}$&1.26/454\\
\hfill& 1.56$^{+0.01}_{-0.01}$ & 5.59$^{+0.15}_{-0.10}$& $<$ 0.21&27$^{+50}_{-12}$&1.26/453\\
\hline
Obs. 7 & 1.52$^{+0.04}_{-0.03}$ & 5.45$^{+0.05}_{-0.05}$&0 (fixed)&27$^{+16}_{-15}$&1.23/372\\
\hfill& 1.52$^{+0.04}_{-0.04}$ & 5.45$^{+0.07}_{-0.06}$&$<$ 0.59 &34$^{+11}_{-24}$&1.23/371\\
\hline
\hline
\end{tabular}
\end{center}
\pn
$^{*}$ The continuum emission was fitted between 2-10 keV with a single absorbed
power law with $N_{\rm H}$=$N_{\rm Hgal}$=1.79 $\times$ 10$^{20}$ cm$^{-2}$.
\pn
\pn
Note: Intervals are at 90 \% confidence for 1 interesting parameter ($\Delta \chi^2$=2.71).
\end{table}

\clearpage

\begin{figure}
\psfig{file=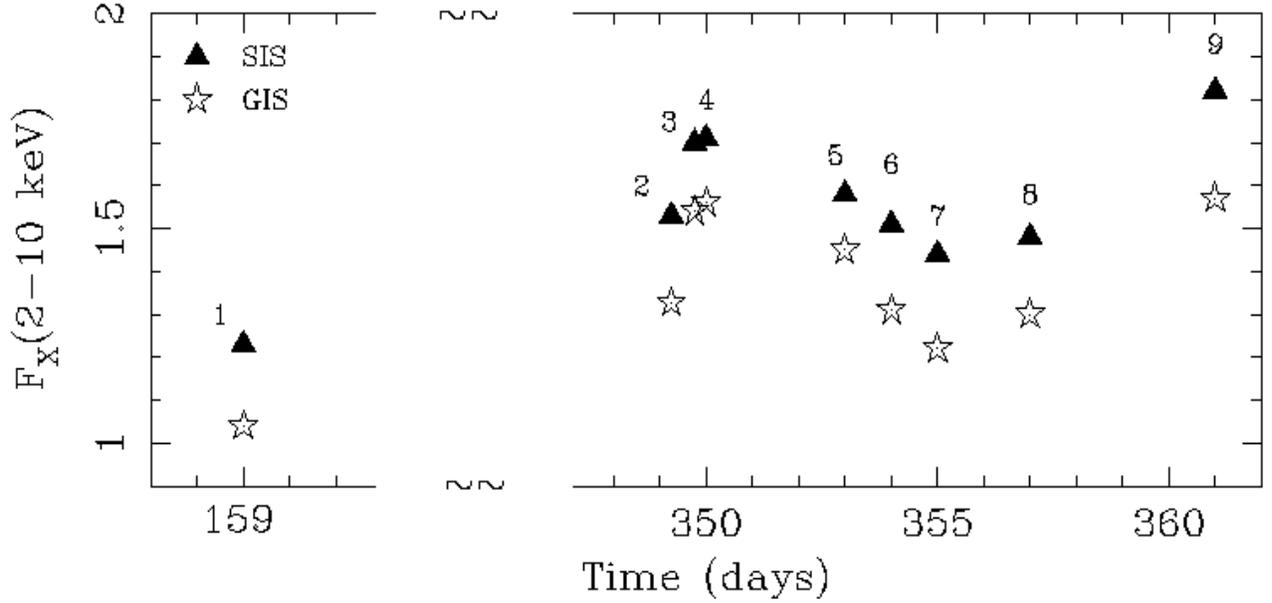,width=18.0cm,height=9.0cm,angle=-90}
\caption{Variation with time (since the mission started) of the 2-10 keV flux.
Note the temporal gap between 
the first and following observations. Fluxes are in units of $10^{-10}$ erg cm$^{-2}$
s$^{-1}$. Flux statistical errors are estimated to be conservatively smaller than 10\%.}
\end{figure}

\begin{figure}
\psfig{file=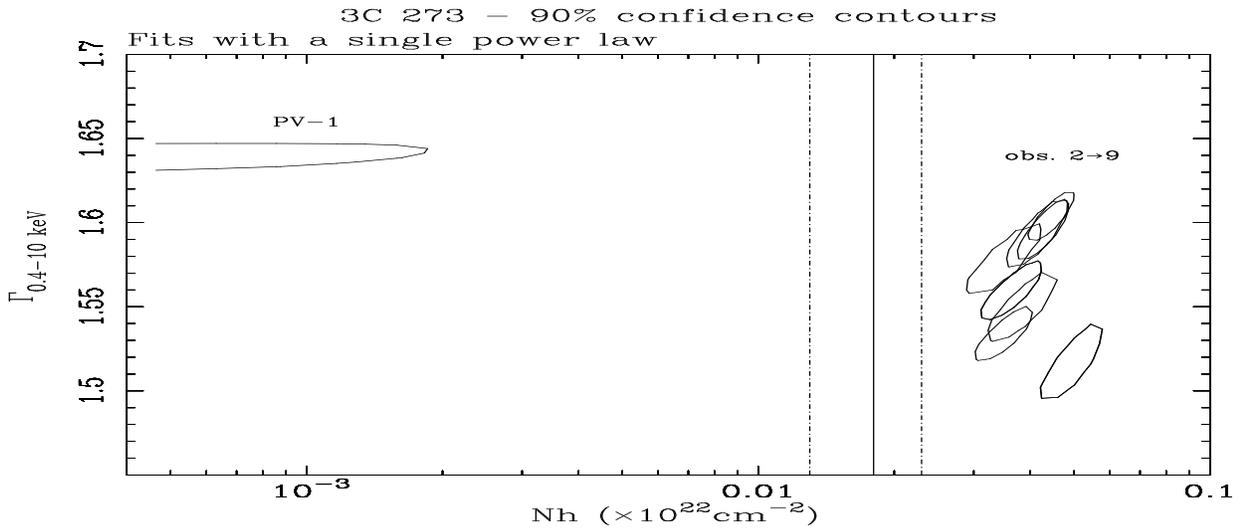,width=18.0cm,height=7.0cm,angle=-90}
\caption{90\% confidence contours of absorption vs photon index obtained
 the SIS, 
for a single absorbed power law model, illustrating the soft-excess, extra absorption
and spectral variability. The vertical line represents the Galactic absorption 
(full line) and associated errors (a conservative error of 30\% as been assumed, 
Dickey \& Lockman 1990).}
\end{figure}

\begin{figure}
\vspace{1cm}
\psfig{file=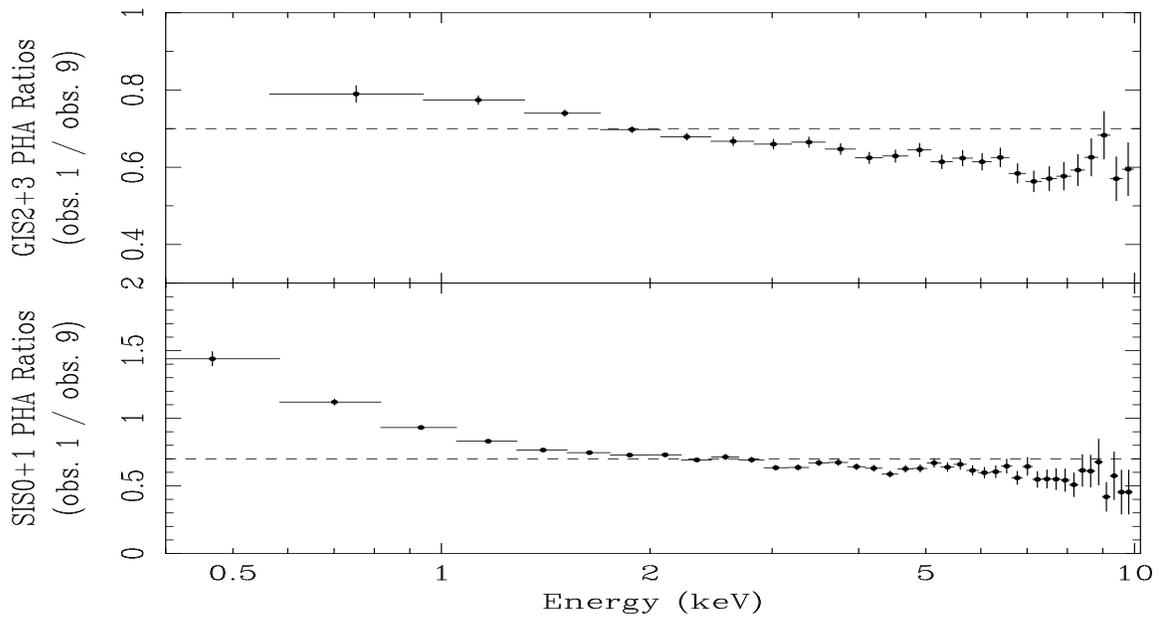,width=18.0cm,height=9.0cm,angle=-90}
\caption{PHA ratios of observation 1 divided by observation 9 which illustrate the 
soft-excess during observation 1. The 2-10 keV spectrum during observation 1 appears 
to be steeper than during observation 9. An horizontal (dashed) line is plotted
to better show the spectral variations. Note that the line-like feature present 
at $\sim$ 8 keV is not statistically significant.}
\end{figure}

\begin{figure}
\psfig{file=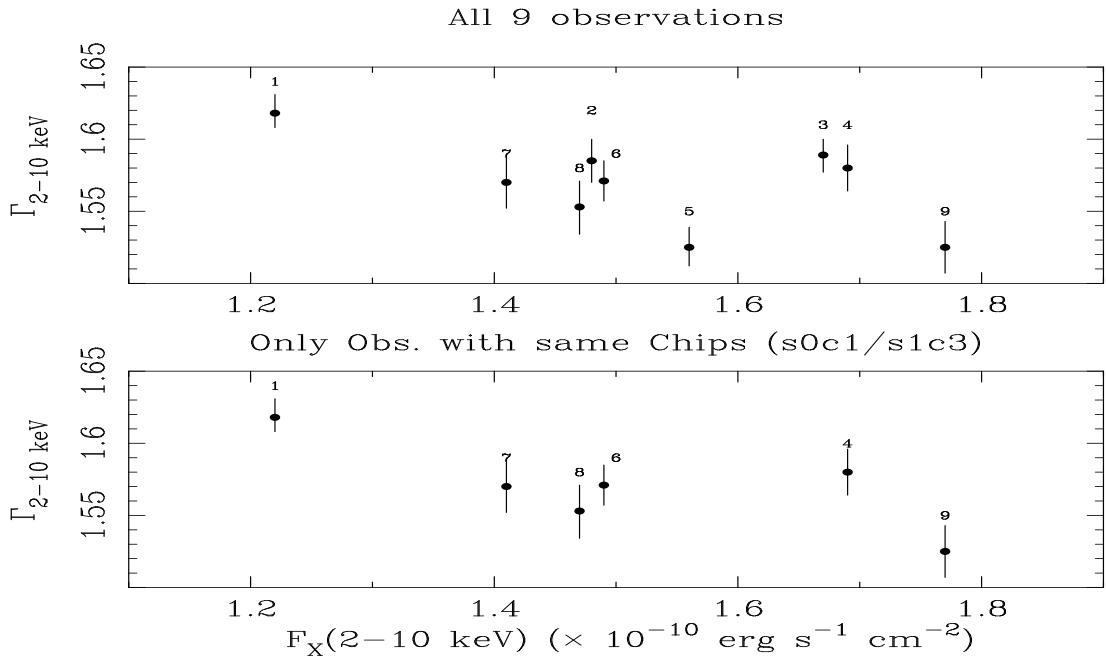,width=18.0cm,height=9.0cm,angle=-90}
\caption{2--10 keV photon index vs. observed 2--10 keV flux. The SIS spectra were fitted 
assuming a power law model with Galactic absorption. Errors are at 90\% confidence for 
one interesting parameter.}
\end{figure}

\begin{figure}
\vspace{-1cm}
\parbox{9truecm}
{\psfig{file=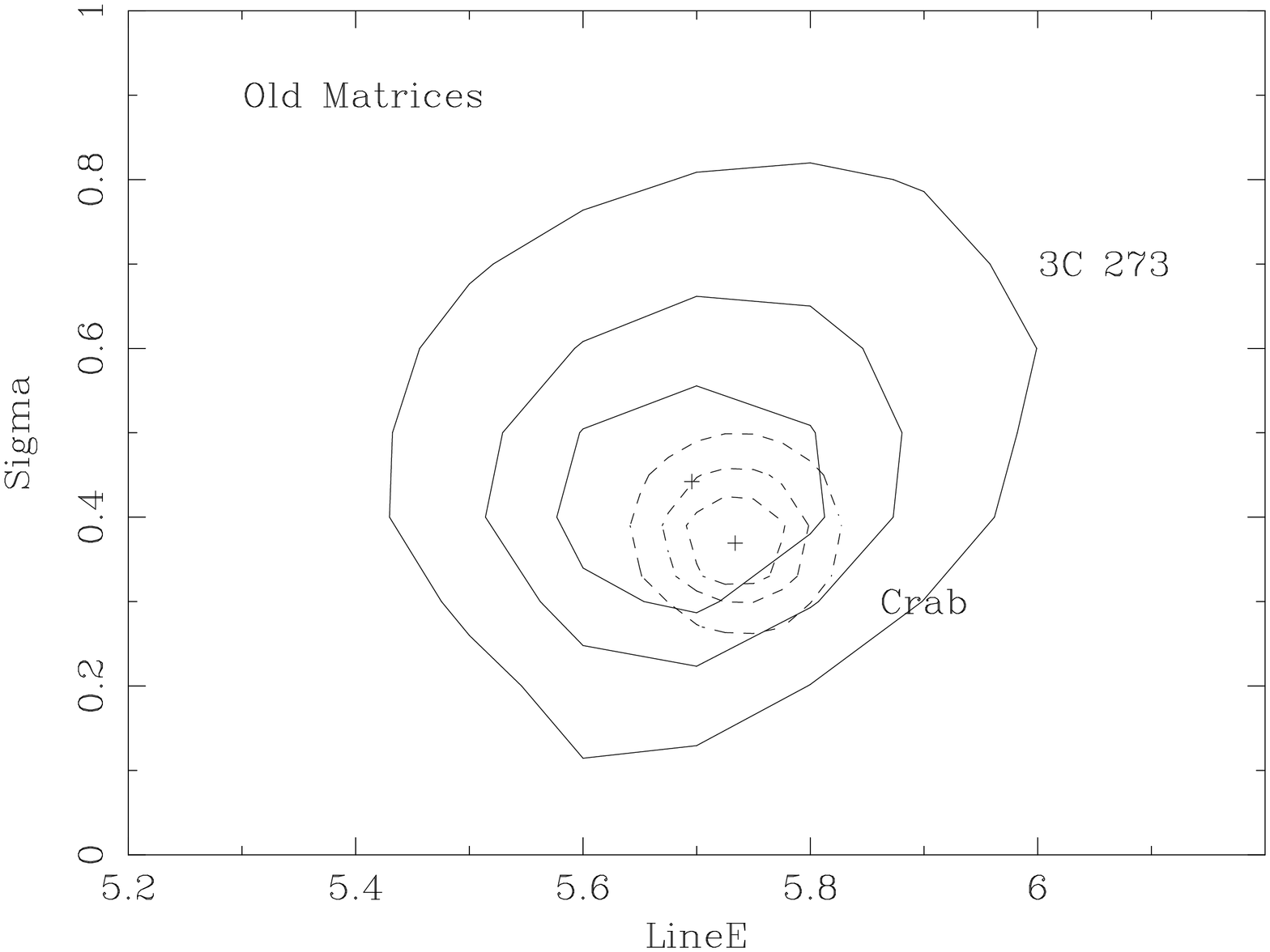,width=8cm,height=8.0cm,angle=-90}}
\  \hspace{0.5truecm}     \
\parbox{9truecm}
{\psfig{file=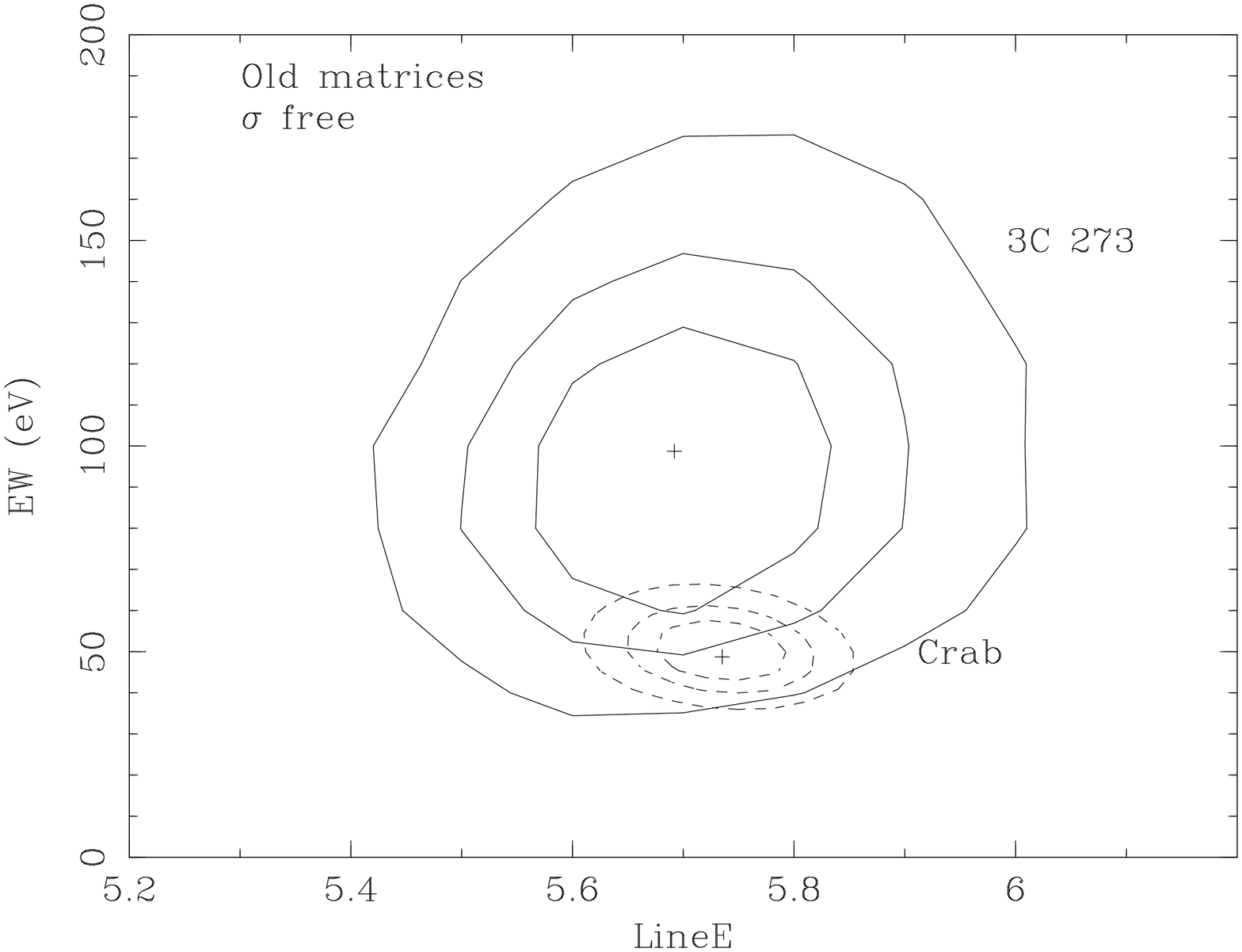,width=8cm,height=8.0cm,angle=-90}}
\pn
\psfig{file=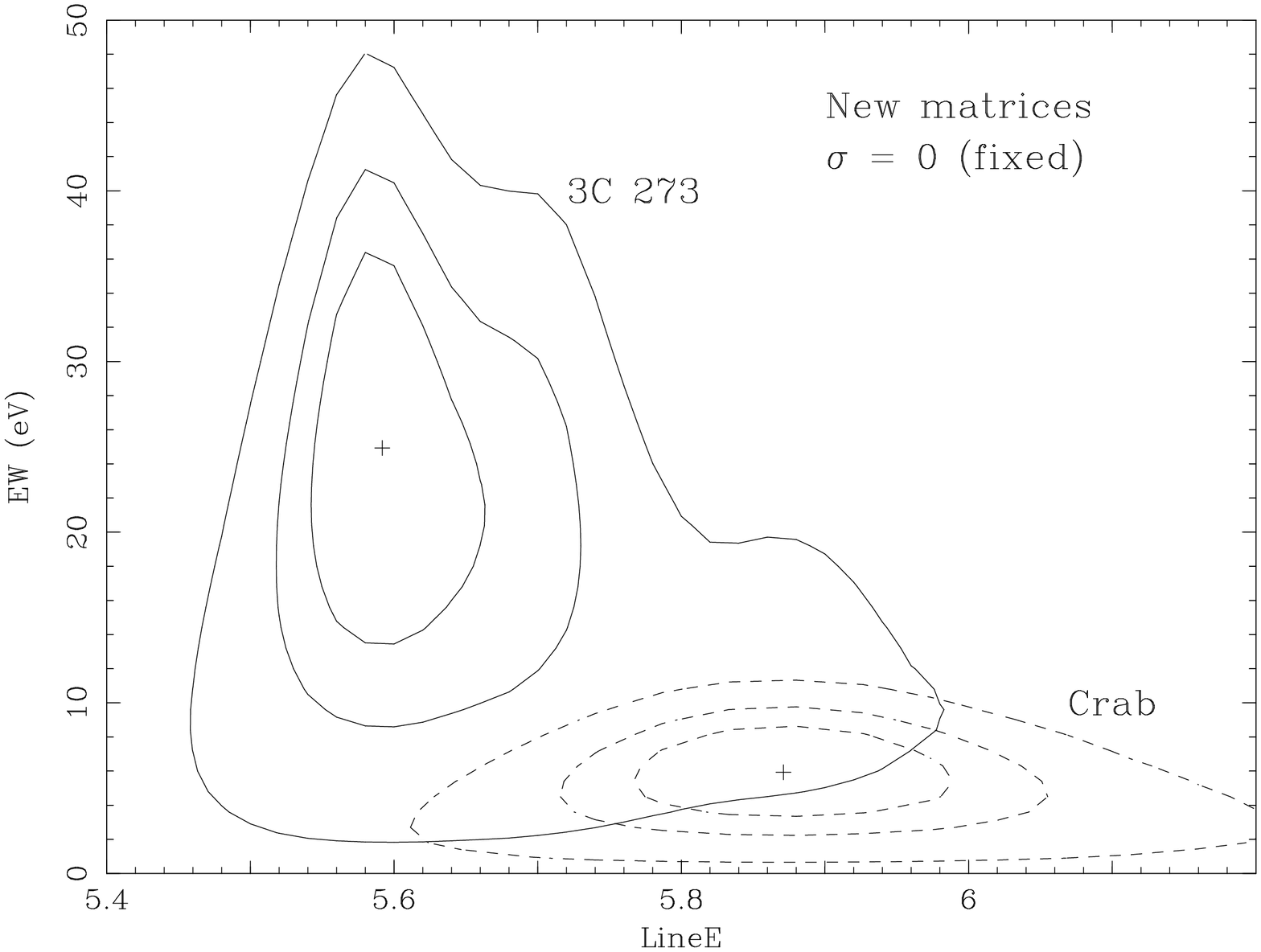,width=18.0cm,height=8.0cm,angle=-90}
\caption{68\%, 90\% and 99\% confidence contour levels of 3C 273 during observation 1
(solid line) and Crab GIS spectrum (dashed line) for the Fe K width vs. energy
(upper left panel) and Fe K equivalent width vs. energy (upper right panel) with
the ``old'' {\it ASCA} calibrations (see text for details).
The lower panel shows the Fe K equivalent width vs. energy contours for the
same observation but with the latest available
calibration. In this case, only upper limits were obtained for the line widths
which were thus fixed to 0 eV. Note the large differences in the best-fit results between
the two calibrations. Note also that the elongated shape of the 99\% confidence line contours
in 3C 273 (lower panel) indicates that, even with the latest calibration matrices
that largely corrected for
the artificial E$\sim$ 5.6--5.8 keV line in the Crab spectrum, it is sensitive to a
very weak (EW$\sim$6-7 eV) line feature apparently still remaining in the Crab data.
This suggests that no more strong calibration uncertainties should affect the measured
line in 3C 273, thus strengthening our conclusion that the line is likely to be real.}
\end{figure}

\begin{figure}
\psfig{file=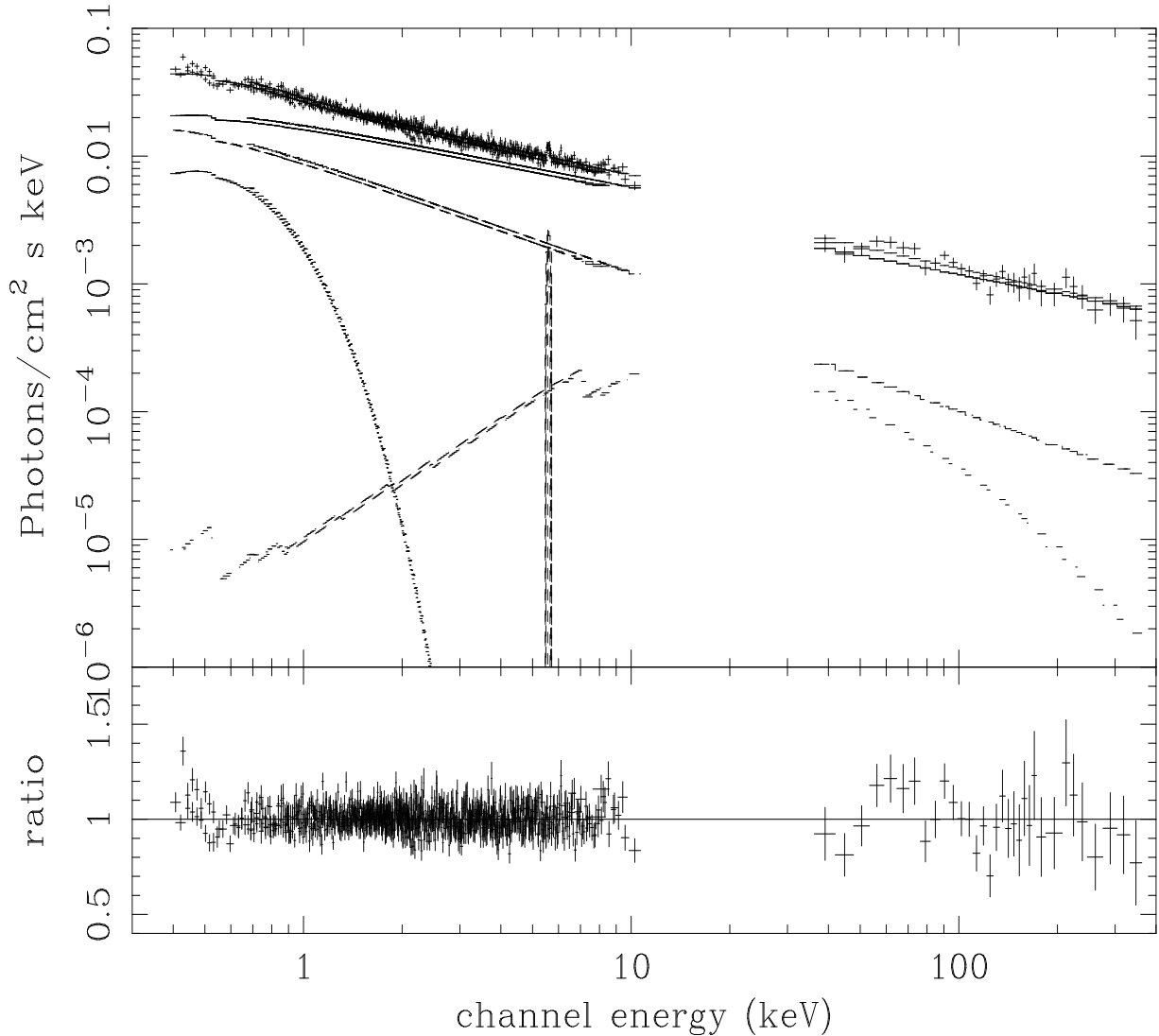,width=18.0cm,height=15.0cm,angle=-90}
\caption{The figure shows the 0.4--300 keV unfolded spectrum and
data/model ratios for a fit of an {\it ASCA} (observation 1) plus {\it OSSE}
(viewing period n.3, Johnson et al. 1995) non-simultaneous data. The figure
{\it illustrates} what could be the various components that contribute to the
X-ray emission in 3C 273, i.e. a $\Gamma \sim$ 1.5-1.6 power law (most likely
due to jet radiation) plus a ``Seyfert-like'' spectrum consisting of a
soft X-ray black-body, a $\Gamma \sim$ 1.9 power law plus a reflection component (with R=1)
and associated iron line (EW=150 eV with respect to the Seyfert continuum). When the
jet continuum is in a low state (like during observations 1 and 7), the features from
the Seyfert contribution are more easily seen. Note that the relative normalizations
of the two instruments were free to vary.}
\end{figure}

\end{document}